# Sustainability and Artificial Intelligence: Necessary, Challenging, and Promising Intersections

Han-Teng Liao
Higher Education Impact Assessment Center
Sun Yat-Sen University Nanfang College
Guangzhou, China

Zijia Wang*
Higher Education Impact Assessment Center
Sun Yat-Sen University Nanfang College
Guangzhou, China
* Corresponding author: wang.zijia@outlook.com

*Abstract*—**Both digital economy and digital technology researchers increasingly recognize the need to better address the role that artificial intelligence (AI) plays in shaping the evolution of the environmental, social and governance aspects of development. It appears that sustainability and AI research converge on the features of wicked problems that are complex, interconnected and dynamic. Building off such convergence, this article aims to map out the necessary, challenging, and promising intersections by providing an overview of the state of art research. Based on 541 bibliographic data collected from the Web of Science (WoS) database, the findings reveal the increasingly central body of work on green and sustainable science and technology in bridging various disciplines, main journals and key topics and concepts. The findings reveal how such interactions can be necessary, challenging, and promising. The article concludes with few general arguments regarding how to diversify and expand the community of practice regarding AI for sustainable development, especially in the areas of expected AI application areas and institutions.**

***Keywords-sustainable development, data revolution, ICT4SD, ICT4D***

## I. INTRODUCTION

Artificial Intelligence (AI)-driven change has created opportunities to improve human well-being and resource management, the main concerns for the notion of sustainability, especially for the developing world which has planet's resources and most of the population[1]. International organizations such as UN[2], UNEP[3], UNESCO[4], etc. have been discussing and implementing AI solutions and applications, as part of the wider Data Revolution or Digital Cooperation strategies, to provide assessment, early warning, and foresight on current and emerging environmental, social and health issues. Also, as part of the digital transformation (i.e. the use of digital technology to transform business, governments, services, etc.), AI can be integrated into solutions as technological, business and societal enablers[5]. When taking sustainability into account, we need AI-enabled green and digital transformation. It appears that sustainability and AI research converge on the features of wicked problems that are complex, interconnected and dynamic. However, to the best of the authors' knowledge, there is little work to examine the intersections of AI and sustainability research. Building off such convergence, this article aims to map out the necessary, challenging, and promising intersections by providing an overview of the state of art research.

The research is funded partly by a project of Smart App Design Innovation Research in the Age of New Business, Arts and Engineering Disciplines (2019GXJK186), under the 2019 Guangdong Education Grants, China, and partly by a curriculum project of "Information Visualization Design" (NFU 02-40250).

## II. SCIENTOMETRIC APPROACH AND DATA

To explore the state-of-art research work at the intersections of sustainability and AI, the article has implemented a query design that covers the notion of AI that includes artificial intelligence, cognitive computing and technologies on the one hand, and the notions of sustainability and sustainable development on the other:

**TS = (("artificial intelligence" OR "cognitive computing" OR "cognitive technolog*") AND ("sustainability" OR "sustainable development") )**

Collected in November 2020, 541 entries of bibliographic data were collected from the Web of Science indexed work in SCI-EXPANDED, SSCI, A&HCI, CPCI-S, CPCI-SSH, ESCI. Python data science packages and VOSviewer were used.

With the aim to identify necessary, challenging and promising intersections, the scientometric approach of the article is designed to explore such interactions at the levels of topics, publication venues, disciplines, institutions and countries. In particular, such scientometric approach allows us to identify the research fronts based on various relationships such as co-cited work clusters, co-word clusters, etc.[6]. By further tracing the evolution of the research fields across time, graphic and tabular presentations are shown so as to examine its dynamics in a more spatial connotation such as disciplines, fields, and geography [7].

## III. RESEARCH FINDINGS

Expected to be recent, cross-disciplinary, and cross-fertilized outcomes, the findings are therefore organized as such.

### A. *Annual trends and piece-wise linear regression modeling*

The overall growth pattern (see the piecewise linear regression modelling outcomes in Fig. 1) shows the year 2009 as the pivotal year after which a much faster growth rate in publications begins. The first time period from 1993 to 2008 contains only 26 articles. The second time period from 2009 to 2020 has seen the growth of publications across journals and disciplines, such as *Sustainability*, *Journal of Cleaner Production*, *IEEE Access*, while *Environmental Modelling & Software*, *Engineering Applications of Artificial Intelligence*, etc. have articles published in both time periods.

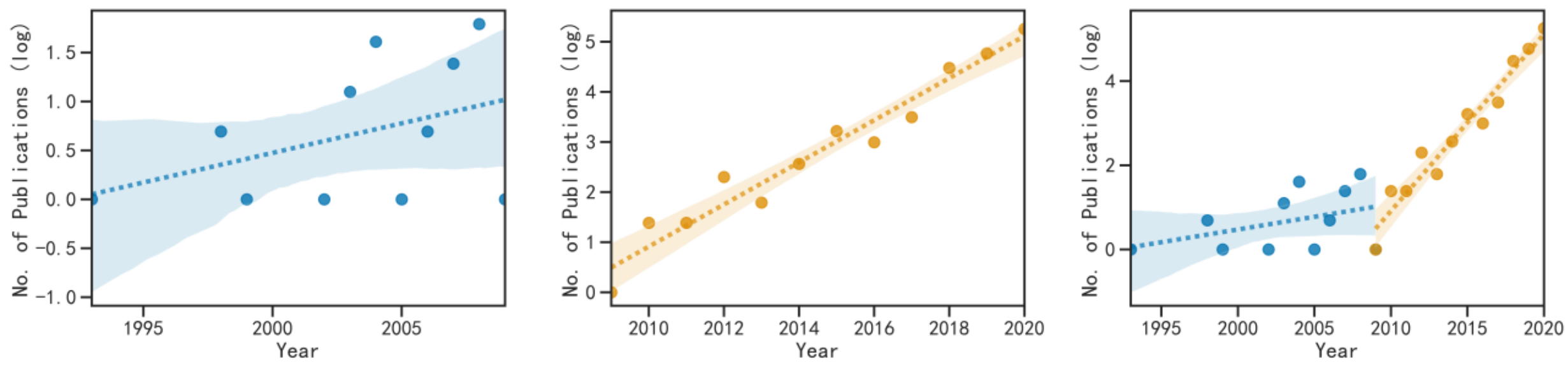

Fig. 1. Trends and piece-wise linear regression modeling: Publications

## B. Main research fronts: publication outlets

Focusing on the 2nd period of publications (N=514), TABLE I reveals the main publication outlets, and their respective disciplines and top topics (based on author keywords). Discipline-wise, as shown in the second column of TABLE I, most journal sources are multi-disciplinary. The numbers in parentheses indicate the ranking numbers: e.g. the top two disciplines, (1-1) and (1-2) have equal number of articles.

Topic-wise, the third column of TABLE I not only shows the attention spread on AI, sustainability, and sustainable development, but also the related key topics across the main journals. While AI appears across almost all selected top journal sources (except *AI & Society* since it already has AI in its journal name), the attention on sustainability and sustainable development is relatively more sporadic. The top 3 journals do cover a much wider range of topics (numbers after the semicolon show the number of articles).

## C. Main research fronts: main disciplines

Altogether, the initial findings of TABLE I indicate three groups of disciplines: (1) environmental sciences and green technology, (2) computer science and AI, and (3) other engineering fields such as chemistry and energy.

Further bibliographic coupling analysis combined with clustering can indicate the tendency of interdisciplinary citation relationship, revealing how main research fronts are composed of different disciplines[8]. Fig. 2 shows the bibliographic coupling network visualization of the 2nd time period at the level of disciplines, defined by the Web of Science categories for each journal [9] (and thus their respective articles).

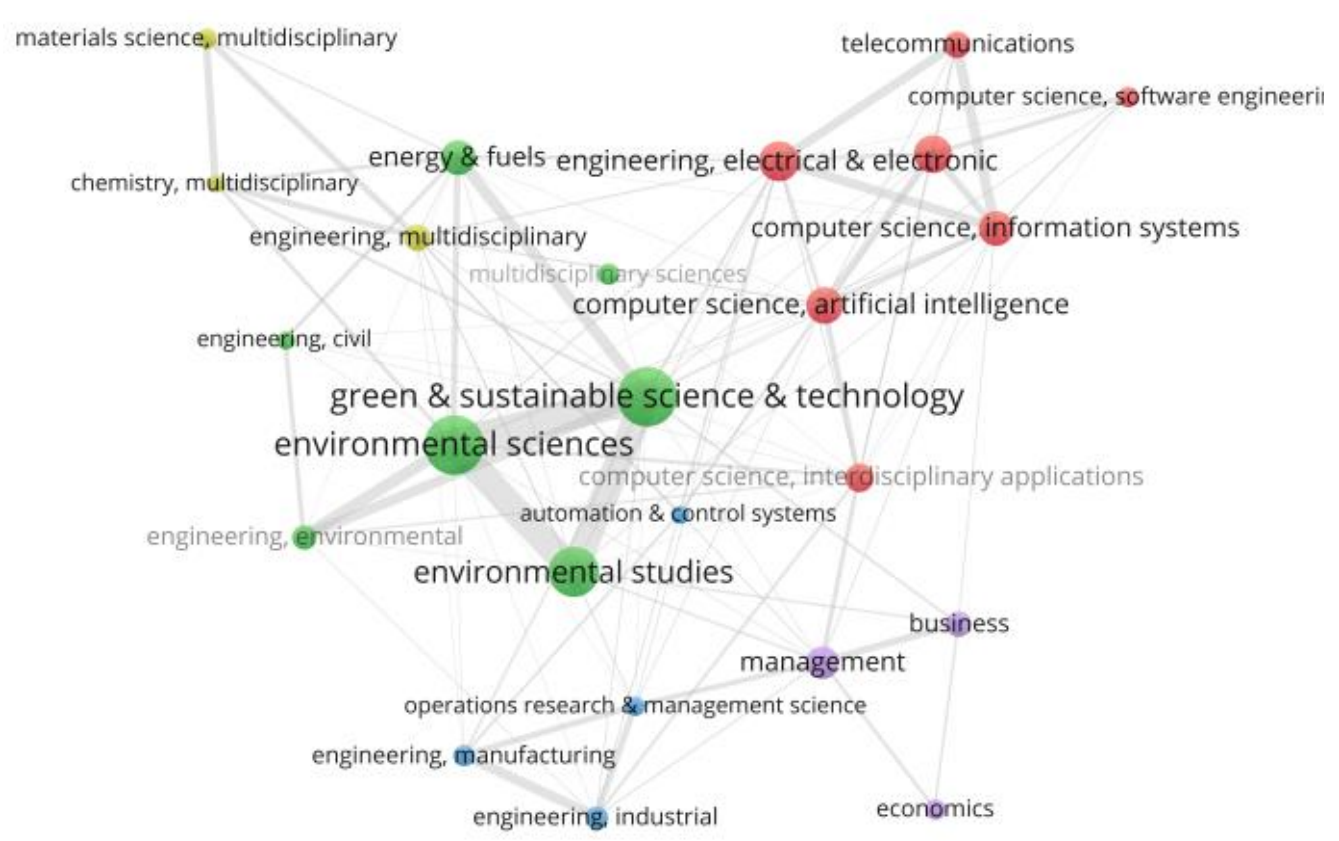

Fig. 2. Main disciplines: a bibliographic coupling network visualization

TABLE I. SELECTED TOP SORCES AND RESPECTIVE DISCIPLINES AND TOPICS

| Sources | *Web of Science Categories (Disciplines)* | *Author Keywords (Topics)* |
|---|---|---|
| (1) Sustainability: 69 | (1-1) Environmental Sciences; (2) Environmental Studies; (1-2) Green & Sustainable Science & Technology | AI: 45; sustainable development: 26; sustainability: 23; healthcare: 7; machine learning: 5; data analytics: 4; IoT: 3; artificial neural network: 3; big data: 3; climate change: 3; data mining: 3; decision making: 3; digital transformation: 3; patent analysis: 3; smart city: 3; support vector machines: 3; bayesian regression: 2; covid-19: 2; genetic algorithm: 2; industry 4.0: 2; resilience: 2; social network analysis: 2; technology: 2 |
| (2) Journal of Cleaner Production: 12 | (11) Engineering, Environmental; (1-1) Environmental Sciences; (1-2) Green & Sustainable Science & Technology | sustainability: 6; AI: 5; sustainable development: 3; IoT: 2 |
| (3) IEEE Access: 9 | (6-1) Computer Science, Information Systems; (3) Engineering, Electrical & Electronic; (9-1) Telecommunications | IoT: 6; AI: 3; healthcare: 3; security: 3; smart city: 3; privacy: 2; sustainability: 2 |
| (4-1) Energies: 6 | (6-2) Energy & Fuels | AI: 8; renewables: 3; smart city: 3; machine learning: 2; smart grid: 2; sustainable development: 2 |
| (5) Applied Sciences-Basel: 5 | (17) Chemistry, Multidisciplinary; (9-2) Engineering, Multidisciplinary; (14-1) Materials Science, Multidisciplinary | AI: 2 |
| (6-2) AI & Society: 4 | (5) Computer Science, Artificial Intelligence | artisanal economy: 2; ethnocomputing: 2; generative justice: 2; human-machine collaboration: 2; industrial symbiosis: 2 |

At the center of Fig. 2, the sustainability discipline cluster consists of mostly sustainable sciences, energy, and civil engineering disciplines. The discipline cluster features journals such as *Sustainability*, *Journal of Cleaner Production*, and *Energies*. At the upper right, the computer science and electrical engineering cluster features journal sources such as *IEEE Access*. At the lower left, the manufacturing and operations cluster features work such as smart manufacturing. At the upper left, the chemistry and material science cluster features work such as data-driven materials discovery [10]. At the lower right, the business and management cluster features work on circular economy, especially in relation to industry 4.0 [11]–[13].

### *D. Main research fronts: key topics*

Revealing the main research fronts in terms of key topics, Fig. 3 shows how the author keywords are clustered. At the center of Fig. 3, the main red cluster contains some of the major issues at the intersections. Shown to be central, the issue of climate change makes the intersections between AI and sustainability necessary. For instance, challenges related to the Internet of Things (IoT) and its sustainability have been outlined [14]. For another, integration of AI, Big Data, etc. into smart cities has been discussed [15], [16], including the disease diagnosis in smart healthcare [17]. Also, largely responding to climate change and the need for sustainable transitions, the ways in which data-driven decision making can solve complex real-world problems have been published in a special issue of *Sustainability*[18]. Such a body of work highlights the importance of smart solutions for the wicked problems towards sustainability.

At the left of Fig. 3, the green cluster includes some of the main promising areas of industry 4.0 and digital economy, using the main concepts such as cyber-physical systems to advance innovations in smart manufacturing and circular economy. Emerging work includes the promising potentials of smart manufacturing[19]–[22], sustainability accounting and reporting [23], largely based on industrial IoT, sensors, and blockchains. At the lower-right of Fig. 3, the blue cluster shows that the combined application of remote sensing, deep learning, artificial neural networks, etc. has contributed to the promising development of precision agriculture [24], Agri-food 4.0[25], nutritional sustainability[26], traceability system [27], etc. At the lower-left of Fig. 3, the yellow cluster reveals the advancement in smart grid and energy[28]–[30]. Finally, at the top of Fig. 3, the blue cluster shows the challenging and promising issues of supply chain management and corporate responsibility, for instance in the applications of blockchains[31], e-commerce [32], etc., for the development of green and sustainable supply chains [33], [34].

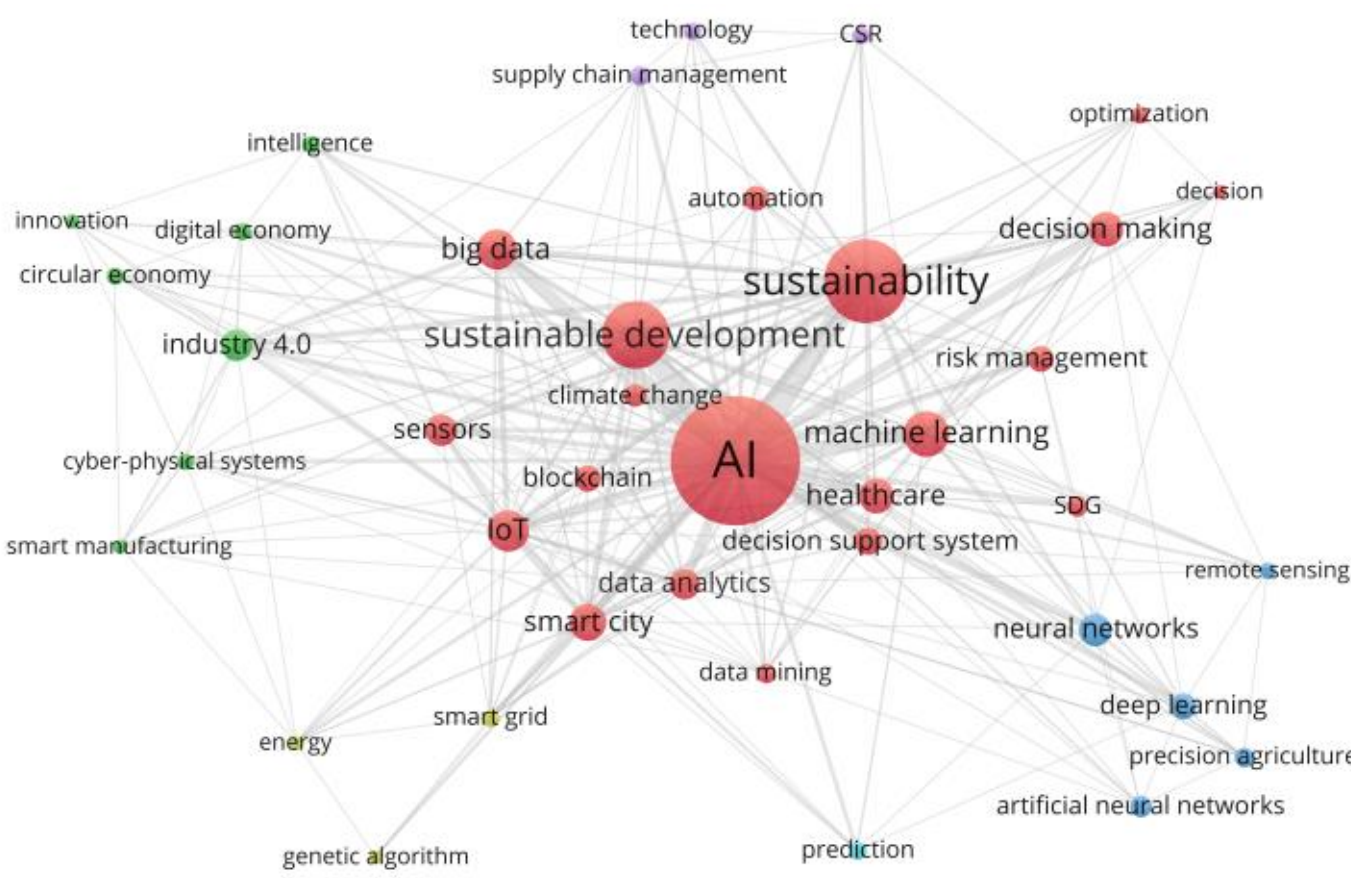

Fig. 3. Top author keywords: a keyword co-occurence network visualization

### *E. Other notable findings*

At the levels of institutions and countries, institutions from China (17%) and the U.S (15%) dominate the intersections, along with other European institutions such as Spain (10%) and England (9%). The top institutions include Vietnamese institutions such as Ton Duc Thang University (8 articles) and Duy Tan University (7 articles), revealing that institutions in the developing countries who seek impacts can make a difference.

## IV. Conclusion

In sum, intersections of sustainability and AI have shown to be necessary, challenging, and promising in various ways. Disciplines such as sustainable sciences, energy, civil engineering, etc. have been leveraging the potentials of AI, Internet of Things, smart grids, etc. to advance the development of smart cities, precision agriculture, healthcare, material discovery, circular economy and industry 4.0. Research work from environmental sciences, green technologies and computer sciences currently dominates the discussions of such intersections, as evidenced by the concentration of publications in the top journals of *Sustainability, Journal of Cleaner Production,* and *IEEE Access*.

Based on the findings, the article concludes with few general remarks on the ways to diversify and expand the community of practice regarding AI for sustainable development, especially in the areas of expected AI application areas and institutions. Researchers from the developing world can play an increasingly significant role. Although the findings presented here are preliminary and limited to the literature using the explicit notions of "sustainability" and "sustainable development", which means the literature mentions “sustainable cities”, “sustainable consumption”, “clean production”, etc. without explicitly uses the terms "sustainability" and "sustainable development" in their bibliographic metadata may have been overlooked. While arguably such omission helps in focusing on the more general implications for sustainability, future work can build upon the findings here to examine a certain aspect of sustainable development in relation to AI.

Overall, between sustainability and AI research, the findings and arguments presented here have discussed the necessary, challenging, and promising intersections, thereby helping the community of practice grow across AI application areas. Future work can continue work on the following questions: How can responsible AI be developed as part of the decision-making systems, corporate social responsibility reporting? How can AI, blockchains, cloud computing and data analytics be designed and integrated for green and digital transformation in areas such as industry 4.0, green supply chain management, sustainable consumption, and green finance?

## Acknowledgment

Han-Teng Liao thanks experts' input from the United Nations University Institute in Macau and participants' input to the keynote speech titled "Big Data and Artificial Intelligence for Sustainable Development Research Fronts" at the 2020 International Conference on Computer Science and Management Technology (ICCSMT2020).

## References

[1] J. Quinn, V. Frias-Martinez, and L. Subramanian, "Computational Sustainability and Artificial Intelligence in the Developing World," *AIMag*, vol. 35, no. 3, pp. 36–47, Sep. 2014, doi: 10.1609/aimag.v35i3.2529.

[2] United Nations General Assembly, *Road map for digital cooperation: implementation of the recommendations of the High-level Panel on Digital Cooperation*. 2020.

[3] UNEP, "Proposed programme of work and budget for the biennium 2020–2021," Nairobi, UNEP/EA.4/4, Mar. 2019.

[4] UNESCO, "Artificial intelligence for Sustainable Development," UNESCO, Paris, France, Synthesis report ED-2019/WS/39, 2019. Accessed: Nov. 25, 2019. [Online]. Available: https://unesdoc.unesco.org/ark:/48223/pf0000370308.

[5] M. Borst, Ed., "Digital Transformation: An IEEE Digital Reality Initiative White Paper." IEEE Digital Reality Initiative, 2020, [Online]. Available: https://digitalreality.ieee.org/images/files/pdf/digital-transformation-white-paper.pdf.

[6] E. Garfield, "Research fronts," *Current Comments*, Oct. 1994, Accessed: Jan. 27, 2020. [Online].

[7] H. Small and E. Garfield, "The geography of science: disciplinary and national mappings," *Journal of Information Science*, vol. 11, no. 4, pp. 147–159, Oct. 1985, doi: 10.1177/016555158501100402.

[8] B. Jarneving, "Bibliographic coupling and its application to research-front and other core documents," *Journal of Informetrics*, vol. 1, no. 4, pp. 287–307, 2007, doi: https://doi.org/10.1016/j.joi.2007.07.004.

[9] M. BOLETTA, "New Web of Science categories reflect ever-evolving research," *Web of Science Core Collection Help*, Jan. 24, 2019. https://clarivate.com/webofsciencegroup/article/new-web-of-science-categories-reflect-ever-evolving-research/ (accessed Nov. 01, 2020).

[10] J. M. Cole, "A Design-to-Device Pipeline for Data-Driven Materials Discovery," *Acc. Chem. Res.*, vol. 53, no. 3, pp. 599–610, Mar. 2020, doi: 10.1021/acs.accounts.9b00470.

[11] S. Rajput and S. P. Singh, "Connecting circular economy and industry 4.0," *International Journal of Information Management*, vol. 49, pp. 98–113, Dec. 2019, doi: 10.1016/j.ijinfomgt.2019.03.002.

[12] T. T. Pham *et al.*, "Industry 4.0 to Accelerate the Circular Economy: A Case Study of Electric Scooter Sharing," *Sustainability*, vol. 11, no. 23, p. 6661, Nov. 2019, doi: 10.3390/su11236661.

[13] M. A. Reuter, "Digitalizing the Circular Economy: Circular Economy Engineering Defined by the Metallurgical Internet of Things," *Metall and Materi Trans B*, vol. 47, no. 6, pp. 3194–3220, Dec. 2016, doi: 10.1007/s11663-016-0735-5.

[14] P. Anand, Y. Singh, A. Selwal, M. Alazab, S. Tanwar, and N. Kumar, "IoT Vulnerability Assessment for Sustainable Computing: Threats, Current Solutions, and Open Challenges," *IEEE Access*, vol. 8, pp. 168825–168853, 2020, doi: 10.1109/ACCESS.2020.3022842.

[15] K. T. Chui, M. D. Lytras, and A. Visvizi, "Energy Sustainability in Smart Cities: Artificial Intelligence, Smart Monitoring, and Optimization of Energy Consumption," *Energies*, vol. 11, no. 11, p. 2869, Oct. 2018, doi: 10.3390/en11112869.

[16] Z. Allam and Z. A. Dhunny, "On big data, artificial intelligence and smart cities," *Cities*, vol. 89, pp. 80–91, Jun. 2019, doi: 10.1016/j.cities.2019.01.032.

[17] K. Chui, W. Alhalabi, S. Pang, P. Pablos, R. Liu, and M. Zhao, "Disease Diagnosis in Smart Healthcare: Innovation, Technologies and Applications," *Sustainability*, vol. 9, no. 12, p. 2309, Dec. 2017, doi: 10.3390/su9122309.

[18] K.-Y. Shen and G.-H. Tzeng, "Advances in Multiple Criteria Decision Making for Sustainability: Modeling and Applications," *Sustainability*, vol. 10, no. 5, p. 1600, May 2018, doi: 10.3390/su10051600.

[19] S. S. Kamble, A. Gunasekaran, A. Ghadge, and R. Raut, "A performance measurement system for industry 4.0 enabled smart manufacturing system in SMMEs- A review and empirical investigation," *International Journal of Production Economics*, vol. 229, p. 107853, Nov. 2020, doi: 10.1016/j.ijpe.2020.107853.

[20] S. Terry *et al.*, "The Influence of Smart Manufacturing towards Energy Conservation: A Review," *Technologies*, vol. 8, no. 2, p. 31, May 2020, doi: 10.3390/technologies8020031.

[21] R. Cioffi, M. Travaglioni, G. Piscitelli, A. Petrillo, and A. Parmentola, "Smart Manufacturing Systems and Applied Industrial Technologies for a Sustainable Industry: A Systematic Literature Review," *Applied Sciences*, vol. 10, no. 8, p. 2897, Apr. 2020, doi: 10.3390/app10082897.

[22] A. Kusiak, "Smart manufacturing," *International Journal of Production Research*, vol. 56, no. 1–2, pp. 508–517, Jan. 2018, doi: 10.1080/00207543.2017.1351644.

[23] K. Tiwari and M. S. Khan, "Sustainability accounting and reporting in the industry 4.0," *Journal of Cleaner Production*, vol. 258, p. 120783, Jun. 2020, doi: 10.1016/j.jclepro.2020.120783.

[24] M. Carolan, "Publicising Food: Big Data, Precision Agriculture, and Co-Experimental Techniques of Addition," *Sociol. Rural.*, vol. 57.0, no. 2, p. 135, Apr. 2017, doi: 10.1111/soru.12120.

[25] M. Lezoche, J. E. Hernandez, M. del M. E. Alemany Díaz, H. Panetto, and J. Kacprzyk, "Agri-food 4.0: A survey of the supply chains and technologies for the future agriculture," *Computers in Industry*, vol. 117, p. 103187, May 2020, doi: 10.1016/j.compind.2020.103187.

[26] S. M. Smetana, S. Bornkessel, and V. Heinz, "A Path From Sustainable Nutrition to Nutritional Sustainability of Complex Food Systems," *Front. Nutr.*, vol. 6, p. 39, Apr. 2019, doi: 10.3389/fnut.2019.00039.

[27] J. Qian *et al.*, "Food traceability system from governmental, corporate, and consumer perspectives in the European Union and China: A comparative review," *Trends in Food Science & Technology*, vol. 99, pp. 402–412, May 2020, doi: 10.1016/j.tifs.2020.03.025.

[28] J.-S. Chou and D.-S. Tran, "Forecasting energy consumption time series using machine learning techniques based on usage patterns of residential householders," *Energy*, vol. 165, pp. 709–726, Dec. 2018, doi: 10.1016/j.energy.2018.09.144.

[29] K. T. Chui, M. D. Lytras, and A. Visvizi, "Energy Sustainability in Smart Cities: Artificial Intelligence, Smart Monitoring, and Optimization of Energy Consumption," *Energies*, vol. 11, no. 11, p. 2869, Oct. 2018, doi: 10.3390/en11112869.

[30] P. D. Diamantoulakis, V. M. Kapinas, and G. K. Karagiannidis, "Big Data Analytics for Dynamic Energy Management in Smart Grids," *Big Data Research*, vol. 2, no. 3, pp. 94–101, Sep. 2015, doi: 10.1016/j.bdr.2015.03.003.

[31] L.-W. Wong, L.-Y. Leong, J.-J. Hew, G. W.-H. Tan, and K.-B. Ooi, "Time to seize the digital evolution: Adoption of blockchain in operations and supply chain management among Malaysian SMEs," *International Journal of Information Management*, vol. 52, p. 101997, Jun. 2020, doi: 10.1016/j.ijinfomgt.2019.08.005.

[32] Duan, Xiu, and Yao, "Multi-Period E-Closed-Loop Supply Chain Network Considering Consumers' Preference for Products and AI-Push," *Sustainability*, vol. 11, no. 17, p. 4571, Aug. 2019, doi: 10.3390/su11174571.

[33] S. Bag, S. Gupta, S. Kumar, and U. Sivarajah, "Role of technological dimensions of green supply chain management practices on firm performance," *JEIM*, vol. ahead-of-print, no. ahead-of-print, Apr. 2020, doi: 10.1108/JEIM-10-2019-0324.

[34] N. R. Sanders, T. Boone, R. Ganeshan, and J. D. Wood, "Sustainable Supply Chains in the Age of AI and Digitization: Research Challenges and Opportunities," *J Bus Logist*, vol. 40, no. 3, pp. 229–240, Sep. 2019, doi: 10.1111/jbl.12224.